\documentclass[prd,floatfix,onecolumn,amsmath,amssymb,10pt]{revtex4-2}
\usepackage{newtxtext,newtxmath} 
\usepackage{graphicx,color,dcolumn,booktabs,bm}
\usepackage{subfigure}
\usepackage{amssymb}
\usepackage{longtable}
\usepackage{indentfirst}
\usepackage{epsfig,gensymb,siunitx}
\usepackage{feynmf}   
\usepackage{epstopdf}   
\usepackage{slashed}  
\usepackage{cases}
\usepackage{xcolor}
\definecolor{navyblue}{RGB}{0,0,150}
\definecolor{CobaltBlue}{rgb}{0,0.28,.67}
\definecolor{maroon}{RGB}{139,25,150}
\usepackage{multirow}
\usepackage{float}
\usepackage{pgf}
\usepackage{graphicx,color,dcolumn,booktabs,bm}
\usepackage[colorlinks, citecolor=purple,anchorcolor=purple,menucolor=purple, linkcolor=purple,filecolor=purple,runcolor=purple,urlcolor=purple,frenchlinks=purple, urlcolor=purple]{hyperref}
\usepackage{physics}
\usepackage{orcidlink}
\usepackage{bbold}
\usepackage[utf8]{inputenc} 
\usepackage{varwidth}

\usepackage{xcolor}

\begin{document}
	\preprint{}
	
	\title{\color{navyblue}{Kaons in  hot and dense QCD}}
	
	\author{K.~Azizi$^{1,2}$}
	\email{kazem.azizi@ut.ac.ir}
	\thanks{Corresponding author}
	\author{G.~Bozkır$^{3}$}
	\author{N.~Er$^{4}$}
	\author{A.~Türkan$^{5}$}

	\affiliation{
		$^{1}$Department of Physics, \href{https://ut.ac.ir/en}{University of Tehran}, North Karegar Avenue, Tehran 14395-547, Iran\\
		$^{2}$Department of Physics, \href{https://www.dogus.edu.tr/en}{Dogus University}, Dudullu-\"{U}mraniye, 34775
		Istanbul,  T\"{u}rkiye}	
		\affiliation{$^{3}$Department of Basic Sciences, \href{https://www.msu.edu.tr/} {Army NCO Vocational HE School, National Defense University}, Altıeylül, 10185 Balıkesir,Türkiye}
		\affiliation{$^{4}$Department of Physics,  \href{https://www.ibu.edu.tr/Website/Default.aspx} {Bolu  Abant İzzet Baysal University}, Gölköy Kampüsü, 14980 Bolu,  Türkiye}
                \affiliation{$^{5}$Department of Basic Sciences,  \href{https://hho.msu.edu.tr/pagedetail.aspx?SayfaId=2&ParentMenuId=2}{Air Force Academy, National Defense University}, Yeşilyurt, 34149, İstanbul, Türkiye}

	\date{\today}

\begin{abstract}

We present a systematic QCD sum-rule analysis of the in-medium properties of the charged  kaon doublet $K^{\pm}$ over the full $(T,\rho)$ plane relevant to current and forthcoming heavy-ion experiments. Working within the QCD sum-rule framework and incorporating temperature- and density-dependent quark, gluon, and mixed condensates, we derive Borel-transformed sum rules for the effective masses $m_{K^{\pm}}$, the pseudoscalar decay constants $f_{K^{\pm}}$, and the vector self-energy $\Sigma_{v}$ of both charged states 
simultaneously. Our vacuum results, $m_{K^{-}} = 494.6^{+4.9}_{-6.9}$~MeV and $f_{K^{-}} = 157.3^{+4.1}_{-2.9}$~MeV (with near-degenerate $K^{+}$ values), are in excellent agreement with Particle Data Group values at the subpercent level. In the medium, $m_{K^{\pm}}$ decreases monotonically with increasing baryon density and temperature, signaling progressive partial restoration of chiral symmetry. A pronounced mass splitting $\Delta m = m_{K^{-}} - m_{K^{+}}$ develops in baryonic matter, driven by the opposite sign of the Weinberg--Tomozawa vector interaction for the two charge states; it reaches $|\Delta m| \sim 0.35$~GeV near $\rho \simeq 3.2\,\rho_{\rm sat}$ at $T = 0$ and is partially quenched by thermal fluctuations. A central outcome of this study is the extraction of the critical onset density $\rho_c$, defined as the threshold beyond which the in-medium modifications of $K^{-}$ properties signal the onset of the transition toward the chirally restored phase. We stress that 
$\rho_c(T)$ should not be interpreted as a precise determination of the QCD critical point—a task beyond the reach of any current effective framework—but rather as an indicator of the 
density regime where hadronic descriptions begin to break down and quark degrees of freedom become increasingly relevant. With this interpretation, our results reveal a pronounced 
temperature dependence: $\rho_c \simeq (1.2$--$1.4)\,\rho_{\rm sat}$ for $T \lesssim 100$~MeV, reflecting the robustness of the hadronic phase under cold compression, 
and a sharp reduction to $\rho_c \simeq 0.45\,\rho_{\rm sat}$ near the pseudo-critical temperature $T_c \simeq 155$~MeV. This dramatic downward shift reflects the synergistic destabilization of the chiral condensate by combined thermal and density fluctuations, and establishes temperature as a more efficient driver of chiral restoration than baryon density alone in the regime accessible to current experiments. Our findings provide quantitative, QCD-based theoretical input for interpreting kaon 
observables at HADES/GSI, CBM/FAIR, STAR/RHIC, and NA61/SHINE/CERN-SPS, and carry direct 
implications for kaon condensation in neutron star matter and the hadronic equation of state 
at supranuclear densities.

\end{abstract}


\maketitle

\section{Introduction} \label{sec:intro}

Quantum Chromodynamics (QCD), the fundamental theory of the strong interaction, governs the dynamics of quarks and gluons. A cornerstone of this theory is chiral symmetry, which is spontaneously broken in the vacuum state. This spontaneous breaking is the mechanism responsible for dynamical quark mass generation and ultimately shapes the observed hadronic spectrum \cite{Hatsuda:1994pi,Fukushima:2010bq,Hayano:2008vn}.

The phase structure of QCD remains a central and challenging area of research in nuclear and particle physics. Under extreme conditions of finite temperature $T$ and/or baryon density $\rho$, such as those generated in relativistic heavy-ion collisions or present inside compact astrophysical objects like neutron stars, the hadronic matter is expected to undergo a transition from a confined hadronic phase to a deconfined state known as the Quark-Gluon Plasma (QGP) \cite{,Fukushima:2010bq,Baym:2017whm}. A key feature of this transition is the partial or complete restoration of chiral symmetry, which manifests through the melting of the chiral condensate $\langle \bar{q}q \rangle$ as the system approaches the critical temperature and/or density \cite{Aoki:2006we,Bazavov:2017dus}.

In the vacuum, the chiral condensate  $\langle \bar{q}q \rangle$ serves as the primary order parameter for chiral symmetry breaking. As the system transitions toward the QGP phase, $\langle \bar{q}q \rangle$ is predicted to decrease significantly, approaching zero in the limit of exact chiral restoration \cite{Lutz:1992uw,Lattimer:2015nhk}. Understanding the behavior of hadrons in hot and dense QCD matter is therefore essential for mapping the phase diagram and identifying the critical boundaries. The in-medium modifications of hadronic properties—particularly masses and decay constants—driven by the reduction of the chiral condensate, provide critical insights into the mechanism of dynamical mass generation and serve as experimental signatures for chiral symmetry restoration \cite{Hatsuda:1994pi,Hayano:2008vn,Mishra:2004te,Song:2018plu,TangHuan2011,Gal:2016boi,Brown1,Cassing:1999}.

Among various hadronic probes, strange mesons, particularly the kaons ($s\bar{q}$ or $\bar{s}q$) serve as exceptional probes of the hot and dense QCD medium \cite{Hatsuda:1994pi,Hayano:2008vn,Mishra:2004te,Song:2018plu,TangHuan2011,Gal:2016boi,Brown1,Cassing:1999,Jurgen,Lutz:1992uw,Lutz:1992dv,Brown1991,Tolos2002,KAPLAN198657}. Since kaons contain a strange (anti-)quark, their properties are highly sensitive to both the light quark condensate $\langle \bar{u}u \rangle + \langle \bar{d}d \rangle$  and the strange quark condensate $\langle \bar{s}s \rangle$, both of which are modified in-medium \cite{Hatsuda:1994pi,TangHuan2011,Costa:2003uu}. The modification of kaon properties—including their effective masses and decay constants—thus provides a direct signature of partial chiral symmetry restoration and offers a window into the flavor structure of the QCD vacuum under extreme conditions \cite{Song:2018plu,Hayano:2008vn,Bozkir:2022lyk,Er:2022cxx}.

Early theoretical works established that pseudoscalar meson properties are tightly connected to the in-medium behavior of quark condensates. Classic studies using chiral perturbation theory \cite{Brown1991}, effective SU(3) chiral models \cite{Mishra:2004te,Gal:2016boi,Brown1,Jurgen,Lutz:1992uw,Lutz:1992dv}, and phenomenological hadronic Lagrangians demonstrated significant changes in kaon masses and interactions in nuclear matter. The pioneering work of Brown and Rho \cite{Brown1991} introduced the concept of in-medium scaling laws, which relate hadron mass modifications to changes in the chiral condensate. QCD sum-rule investigations \cite{Hatsuda:1994pi,Song:2018plu,Hayano:2008vn,Kumar:2013tna,Cohen:1994wm} further quantified kaon mass shifts and decay-constant suppression at finite density by incorporating medium-dependent condensates into the operator product expansion (OPE) framework. More recent analyses have extended these methods to finite temperature \cite{TangHuan2011,Costa:2003uu,Chanfray:2000ih,Cabrera:2008tja,Ilner:2013ksa,Rapp:2009yu,Kumar2022,Cassing:2015owa,Ahmad:2024ohm,Kahangirwe:2024cny,Costa:2019bua,Moreau:2017,Bozkir:2022lyk}, highlighting the role of thermal condensate evolution and the approach to the chiral transition.

Conventionally, the high-temperature QCD transition was viewed as a single, rapid crossover around $T_c \sim 155$ MeV where chiral symmetry restoration $T_{\textrm{ch}}$ and deconfinement occurred nearly simultaneously \cite{Aoki:2006we,Bazavov:2017dus}. However, recent lattice QCD simulations \cite{Bazavov:2017dus,Ding:2015ona,Borsanyi:2020fev}  and theoretical studies \cite{Glozman:2022zpy} suggest a more nuanced picture, pointing to an  $T_{\textrm{ch}}$ intermediate regime above —termed the 'Stringy Fluid' or 'semi-QGP'—that may persist up to approximately $3\,T_{\textrm{ch}}$. This regime is characterized by partial deconfinement with remnants of hadronic correlations, emphasizing the complexity of the QCD phase structure.

A consistent and striking feature emerging from these studies is the strong asymmetry between the in-medium behavior of $K^+$ and $K^-$ mesons in baryonic matter, which arises fundamentally from the opposite sign of the Weinberg–Tomozawa vector interaction term in $KN$ scattering \cite{Mishra:2004te,Song:2018plu,TangHuan2011,Jurgen,Lutz:1992uw,Lutz:1992dv}. This leads to a repulsive vector potential for  $K^+$ and an attractive potential for $K^-$ in dense baryonic medium, causing a mass splitting that grows systematically with baryon density—a result robustly supported by chiral SU(3) effective models \cite{Mishra:2004te,Song:2018plu,TangHuan2011,Gal:2016boi,Jurgen,Lutz:1992uw,Lutz:1992dv}, coupled-channel treatments \cite{Ilner:2013ksa}, and QCD sum-rule calculations \cite{Song:2018plu,Hayano:2008vn,Bozkir:2022lyk,Er:2022cxx}. At finite temperature and low baryon density, the dominant modification instead originates from the thermal reduction of both the strange and light quark condensates \cite{TangHuan2011,Costa:2003uu,Costa:2019bua}, producing a comparable suppression trend for both $K^+$ and $K^-$ masses and decay constants.

The phenomenological implications of these in-medium modifications are profound and multifaceted. Experimental analyses in relativistic heavy-ion collisions, including collective flow patterns\cite{Cassing:1999,Cassing:2015owa,Moreau:2017}, yield ratios \cite{Brown1,Cassing:1999}, and strangeness production rates \cite{Mishra:2004te,Brown1,Kornakov:2018kxg,HADES:2018qkj,CBM:2016kpk,STAR:2020dav,NA61SHINE:2020czq}, exhibit sensitivity to the density- and temperature-dependent effective masses and interaction cross-sections of kaons. The HADES collaboration at GSI \cite{Kornakov:2018kxg,HADES:2018qkj} and the NA61/SHINE experiment at CERN SPS \cite{NA61SHINE:2020czq} have provided crucial data on kaon production and absorption in dense hadronic matter, while the upcoming Compressed Baryonic Matter (CBM) experiment at FAIR \cite{CBM:2016kpk} aims to systematically explore the high-density regime of the QCD phase diagram. Furthermore, in neutron-rich dense matter relevant for neutron star interiors and proto-neutron star evolution, the possibility of $K^-$  condensation—a phenomenon that could dramatically soften the equation of state and affect the maximum mass of neutron stars—depends crucially on the in-medium mass and decay constant of the $K^-$ meson \cite{Baym:2017whm,KapNelson,Alford:1997zt,Tolos2002,KAPLAN198657,Lattimer:2015nhk}.

Understanding the QCD phase diagram at high baryon densities requires a detailed and systematic analysis of how the nuclear medium affects fundamental hadronic observables such as masses, decay constants, and interaction vertices. As highlighted by experimental programs at GSI-HADES \cite{Kornakov:2018kxg,HADES:2018qkj} and the future FAIR facility \cite{CBM:2016kpk}, probing the properties of hot and dense matter is essential for identifying robust signatures of chiral symmetry restoration and for constraining the hadronic equation of state at supra-nuclear densities. Our current study builds upon this motivation by providing a comprehensive theoretical evaluation of the $K^{\pm}$ meson mass and decay constant shifts within the QCD sum rule framework, which are central to interpreting experimental results and advancing our understanding of the strong interaction under extreme conditions.

In this context, the present work provides a systematic theoretical analysis of the strange pseudoscalar meson sector in hot and dense QCD matter. We calculate the effective masses $m_{K^{\pm}}$ and decay constants $f_{K^{\pm}}$ for both the $K^+$ and $K^-$ mesons as functions of temperature ($T$) and baryon density ($\rho$) within the well-established QCD sum rules (QCDSRS) framework. The QCDSRS method, which relies on the operator product expansion and incorporates medium-modified quark and gluon condensates, provides a rigorous non-perturbative approach to extract hadronic properties from fundamental QCD degrees of freedom. Crucially, we also determine the mass splitting $\Delta m_{K^{\pm}} = m_{K^{-}} -m_{K^{+}}$ over the entire $(T, \rho)$
 plane accessible to current and future heavy-ion experiments.
 
 Our results aim to quantify systematically the effects of partial chiral symmetry restoration and finite baryon density on these fundamental observables, providing valuable theoretical input for interpreting experimental data from HADES  \cite{Kornakov:2018kxg,HADES:2018qkj}, CBM at FAIR \cite{CBM:2016kpk} STAR at RHIC \cite{STAR:2020dav}, and NA61/SHINE at CERN SPS \cite{NA61SHINE:2020czq}. By elucidating the microscopic mechanisms governing kaon property modifications—rooted in the density and temperature dependence of QCD condensates—our study offers a unified picture of strangeness dynamics in dense matter, with direct implications for both relativistic heavy-ion collision phenomenology and the equation of state of compact astrophysical objects \cite{Baym:2017whm,Lattimer:2015nhk,Fukushima:2013rx}.
 
 The determination of hadronic masses and decay constants in vacuum and in-medium is effectively achieved using QCDSRS \cite{Hatsuda:1994pi,Cohen:1994wm}, a remarkably successful non-perturbative method that bridges the gap between the underlying quark-gluon dynamics and observable hadronic properties. The QCD vacuum is characterized by several fundamental condensates, notably the chiral (quark) condensate  $\langle \bar{q}q \rangle$, the gluon condensate $G^2_{\mu\nu}$, and the mixed quark-gluon condensate $\langle \bar{q}g_s \sigma G q \rangle$. Extending these calculations to dense and hot mediums necessitates the systematic incorporation of temperature- and density-modified condensates, which encode the response of the QCD vacuum to the presence of finite baryon density and thermal excitations \cite{Song:2018plu,TangHuan2011,Hayano:2008vn,Costa:2003uu,Bozkir:2022lyk,Er:2022cxx,Kumar:2013tna}. This approach has been successfully applied to various mesonic and baryonic systems \cite{Song:2018plu,Hayano:2008vn,Kumar2022,Bozkir:2022lyk,Er:2022cxx,Kumar:2013tna,Hatsuda:1994pi,Cohen:1994wm}, yielding predictions that are consistent with experimental observations and lattice QCD simulations where available \cite{Aoki:2006we,Bazavov:2017dus,Ding:2015ona,Borsanyi:2020fev}.
 
 The rest of this paper is organized as follows: In Sec. II, we derive the QCD sum rules for the masses and decay constants of the $K^+$ and $K^-$ mesons at finite temperature and baryon density, incorporating the temperature- and density-dependent modifications of the relevant QCD condensates. Section III presents and discusses the numerical results obtained from these sum rules, including the behavior of $m_{K^{\pm}}$, $f_{K^{\pm}}$, and the mass splitting $\Delta m_{K^-,K^+}$ as functions of $T$ and $\rho$, as well as the determination of the critical onset point $\rho_c$. Finally, Sec. IV provides concluding remarks and discusses the implications of our findings for heavy-ion collision experiments and compact star physics.
 

\section{QCDSRS In Hot and Dense Medium} \label{sec:1}

The time-ordered two-point correlation function $\Pi_{\mu\nu}$, relevant to the calculation of the kaon mass and decay constant in a hot and dense medium, is given by:
\begin{equation}
\label{pimunu}
\Pi_{\mu\nu}(T,\rho) = i \int d^4 x e^{ip\cdot x} \langle\Omega_0|\mathcal{\textbf{T}}[\mathcal{\textbf{J}}^{K^{\pm}}_{\mu}(x)\mathcal{\textbf{J}}^{K^{\pm}\dagger}_{\nu}(0)]|\Omega_0\rangle,
\end{equation}
where $|\Omega_0\rangle$ represents the ground state of the hot and dense medium. For vacuum calculations, this state is replaced by the vacuum state $|0\rangle$.  The symbol $\mathcal{\textbf{T}}$ stands for the time-ordered product of the two currents. The interpolating currents for the $K^+$ and $K^-$ mesons used in Eq. (\ref{pimunu}) are given by:

\begin{equation}
\label{current}
\mathcal{\textbf{J}}^{K^{\pm}}_{\mu}(x)=\bar{q_1}^a(x)\gamma_{\mu}\gamma_{5} q_2^a(x),
\end{equation}
where summation over the color index is implied, and $\gamma_{\mu}$ and $\gamma_{5}$ are the standard Dirac matrices. The quark fields are specified as $(q_1, q_2) = (s, u)$ for the $K^+$ meson, and $(q_1, q_2) = (u, s)$ for the $K^-$ meson.

The evaluation of the correlation function  Eq. (\ref{pimunu}) exploits the duality between two complementary representations. The first, the hadronic (or physical) side, is expressed in terms of hadronic properties. The second, the QCD (or theoretical) side, is expressed in terms of fundamental QCD parameters such as quark masses and medium-dependent condensates. By matching the coefficients of corresponding tensor structures on both sides, QCDSRS are established for the observables of interest. Following the calculation of the correlation function in $x$-space and its transformation to momentum space, a Borel transformation is performed on the resulting equation. This step effectively suppresses the contributions from higher states and the continuum. The final stage involves applying a continuum subtraction procedure under the assumption of quark-hadron duality.


\subsection{Hadronic framework}
The interpolating current for the kaon defined in Eq. (\ref{current}) couples to both the pseudoscalar $(PS)$ and axial vector $(AV)$ kaon channels. To derive the hadronic representation of the correlator, we insert a complete set of intermediate hadronic states between the interpolating currents in Eq. (\ref{pimunu}). After performing the four-dimensional integral over the space-time variable $x$, the hadronic representation is expressed in terms of the kaon's effective mass ($m^{*}_{K^{\pm}}$) and decay constant ($f^{*}_{K^{\pm}}$) in the hot and dense medium. Since the axial current interacts with both PS and AV kaons, isolating these two distinct contributions is necessary after the spacetime integration in order to obtain the final sum rule. Thus, the hadronic representation reads:

\begin{equation}
\label{corre3}
\Pi^{Had}_{\mu\nu}(T,\rho)  =\frac{ \langle\Omega_0|\mathcal{\textbf{J}}^{K^{\pm}}_{\mu}|K^{\pm}_{AV}(p^*)\rangle  \langle K^{\pm}_{AV} (p^*)|\mathcal{\textbf{J}}^{K^{\pm}\dagger}_{\nu}|\Omega_0\rangle}{m_{K^{\pm},AV}^{*2}-p^{*2}} +\frac{ \langle\Omega_0|\mathcal{\textbf{J}}^{K^{\pm}}_{\mu}|K^{\pm}_{PS}(p^*)\rangle  \langle K^{\pm}_{PS} (p^*)|\mathcal{\textbf{J}}^{K^{\pm}\dagger}_{\nu}|\Omega_0\rangle}{m_{K^{\pm},PS}^{*2}-p^{*2}} + (...).
\end{equation}
The ellipsis ($\dots$) in this expression represents the contributions arising from higher resonances and continuum states in both the $AV$ and $PS$ kaon channels. In a hot and dense medium, the effective mass is a function of the baryon density ($\rho$) and temperature ($T$), i.e., $m^* = m(T, \rho)$. The four-momentum is modified to $p^*_{\mu}=p_{\mu}-\Sigma_{\upsilon }u_{\mu}$, where, $\Sigma_{\upsilon }$ is the scalar coefficient of the vector self-energy, which in general can be decomposed as
\begin{equation}
\Sigma^{\mu}_{\upsilon} = \Sigma_{\upsilon} u^{\mu} + \Sigma'_{\upsilon} p^{\mu},
\end{equation}
where $ \Sigma_{\upsilon}$ and $\Sigma'_{\upsilon}$ are Lorentz scalar functions and $u_{\mu}$ is the four-velocity of the medium. Here we neglect  $\Sigma'_{\upsilon}$ due to its small
contribution. Henceforth, we adopt the rest frame of the medium, where  $u_{\mu}=(1,0)$. For convenience, we express the hadronic matrix elements in terms of the kaon's effective mass, decay constant, and momentum within the medium, incorporating the AV state's polarization vector:
\begin{align}
\label{defff}
  \langle\Omega_0|\mathcal{\textbf{J}}^{K^{\pm}}_{\mu}|K^{\pm}_{AV}(p^*)\rangle  &= f^*_{AV} m^*_{AV} \zeta^*_{\mu}, \nonumber \\
 \langle\Omega_0|\mathcal{\textbf{J}}^{K^{\pm}}_{\mu}|K^{\pm}_{PS}(p^*)\rangle  &= i f^*_{PS} p^*_{\mu},
\end{align}
where $f^*_{AV}$ and $f^*_{PS}$ denote the temperature- and density-dependent decay constants of the kaon in the $AV$ and $PS$ channels, respectively. The polarization vector $\zeta_\mu$ for the $AV$ state satisfies the following condition:
\begin{equation}
\sum \zeta_{\mu} \zeta^*_{\nu}=\Big[-g_{\mu\nu}+\frac{p_{\mu}^{*}p_{\nu}^{*}}{p^{*2}}\Big].
\end{equation}

To isolate the $PS$ contribution, the $AV$ contributions must be subtracted. In medium, we project onto the $p^*_{\mu}$ structure to achieve this separation. Beyond the vacuum case, a richer set of independent tensor structures and additional constraints are required to determine parameters such as the vector self-energy. Constructing a system of independent sum rules allows one to simultaneously extract the effective mass ($m^*$), the decay constant ($f^*$), and the vector self-energy ($\Sigma_{\upsilon} $) of the kaon. Furthermore, to suppress the influence of higher resonances and the continuum, a Borel transformation in $p^2$ is applied. Following the removal of AV contributions and the implementation of the standard QCDSRS procedure, the Borel-transformed hadronic side for the
hot and dense medium reads:

\begin{equation}
\mathbf{ \hat{B}}\big[p^*_{\mu}\Pi^{Had}_{\mu\nu}\big](T,\rho)  = - f_{PS}^{*2}m_{PS}^{*2} e^{-\mu^2/M^2} (p_{\nu}-\Sigma_{\upsilon} u_{\nu}),\nonumber\\
\end{equation}
here, we define $\mu^2 = m_{PS}^{*2}-\Sigma^2_{\upsilon} +2 \Sigma_{\upsilon} p_0$. The variable $M^2$ denotes the Borel mass scale, $p_0$ refers to the in-medium quasi-particle energy.


\subsection{QCD framework}

Alternatively, the correlation function defined in Eq. (\ref{pimunu}) can be evaluated by considering the quark and gluon degrees of freedom within the deep Euclidean region. By inserting the interpolating current from Eq. (\ref{current}) into the correlation function and performing the necessary contractions of the light quark fields, the QCD (theoretical) side is derived as follows:

\begin{eqnarray}
\label{corre1}
\Pi^{QCD}_{\mu\nu} (T,\rho)
=
\left\{\begin{array}{l}i \int d^4 x\, e^{ip\cdot x}\,\mathrm{Tr}\!\left[\gamma_5 \mathcal{G}^{mn}_{s}(x)\gamma_5 \gamma_{\nu}\mathcal{G}^{nm}_{u}(-x)\gamma_{\mu}
\right] \quad \text{for } K^-,
\\[2mm]i \int d^4 x\, e^{ip\cdot x}\,\mathrm{Tr}\!\left[\gamma_5 \mathcal{G}^{mn}_{u}(x)\gamma_5 \gamma_{\nu}\mathcal{G}^{nm}_{s}(-x)\gamma_{\mu}\right] \quad \text{for } K^+.
\end{array}
\right.
\end{eqnarray}
where $ \mathcal{G}^{mn}_{q}$ (with $q=u, s$) denotes the full color-singlet light quark propagators in the hot and dense medium. In the deep Euclidean region, the light quark propagator $\mathcal{G}^{mn}_q(x)$ admits an Operator Product Expansion of the following form:
\begin{eqnarray}\label{FDF}
\label{prop}
\mathcal{G}^{mn}_q(x) &=& (1-\mathcal{F})\Big(\frac{i}{2\pi^2}\frac{\slashed{x}}{ x^4}-\frac{m_q}{4\pi^2x^2}+i\frac{m_q^{2} }{8\pi^2 } \frac{\slashed{x}}{x^2}\Big)\delta^{mn} + \chi_q^m(x) \bar{\chi}_q^n(0)  - i\frac{g_s}{32 \pi^2}\frac{\slashed{x}\sigma_{\mu \nu}+\sigma_{\mu \nu}\slashed{x}}{x^2}F^{\mu \nu}_A (0) t_A^{mn}\nonumber\\
&+&\frac{i}{3}\Big[\Big(-\frac{\slashed{x}}{12} \langle u^{\mu} \Theta^f_{\mu\nu}u^{\nu}\rangle\Big) + \frac{1}{3}\Big(u\cdot x \slashed{u}\langle u^{\mu} \Theta^f_{\mu\nu}u^{\nu}\rangle\Big)\Big]\delta^{mn} + ...,
\end{eqnarray}
here, the thermal effects are incorporated through the Fermi-Dirac function in perturbative part, $\mathcal{F}=(e^{\beta/ |x_0|}+1)^{-1}$, where $\beta$ is the inverse temperature. The factor $(1-\mathcal{F})$ accounts for the thermal modification of the quark propagator in a hot medium, with its detailed derivation provided in the Appendix A. Physically, it reflects the Fermi-Dirac statistics of quarks and the resulting Pauli blocking effect, ensuring that the propagator reduces to its vacuum form as $T \to 0$ (or $\beta \to \infty$).

This factor is applied specifically to the perturbative components of the propagator to account for the statistical occupancy of the thermal bath. Since the perturbative part describes the propagation of quasi-free quarks, it must be weighted by the probability of finding an empty state, consistent with the Pauli Exclusion Principle. 

In contrast, the non-perturbative terms (condensates) represent background field configurations whose temperature and density dependencies are already intrinsically captured by their in-medium expectation values. These include the quark, gluon, and mixed condensates within the hot and dense medium. A comprehensive treatment of the non-perturbative contributions to the propagator in thermal and dense regimes is detailed in Refs. \cite{Turkan:2021dvu} and \cite{Er:2022cxx}, respectively. Nevertheless, to examine the combined influence of thermal and medium effects on the physical observables, it has become necessary to update the non-perturbative condensates within the current framework.

In order to capture the profile of the chiral condensate at finite temperature and density,  we parametrize it by fitting the data digitized from Ref. \cite{TangHuan2011}, which leads to the following functional form:
\begin{align}
 \langle  \bar{q}q\rangle_{(T, \rho)} =&   \langle  \bar{q}q\rangle_0 \Big[ 1.0-0.0389 \,  (\rho/\rho_{\text{sat}}) +0.0007\, (\rho/\rho_{\text{sat}})^2 \nonumber\\
 &+ 2.2014 \, T+0.0218 \,  (\rho/\rho_{\text{sat}})^2 \, T -20.7720 \, T^2  - 8.2014 \, (\rho/\rho_{\text{sat}}) \, T^2  \Big]  \, \text{GeV}^3,
\end{align}
where $\rho_{\text{sat}}$ is  the nuclear matter saturation density, $\rho_{\text{sat}} = 0.11^3 \text{ GeV}^3$ and all coefficients are determined from the fit. The vacuum chiral condensate is taken as $\langle \bar{q}q \rangle_0 = (-0.241)^3 \text{ GeV}^3$. Furthermore, the strange quark condensate at finite temperature and density is assumed to follow the relation $\langle \bar{s}s\rangle_{(T,\rho)} = 0.8 \langle \bar{q}q\rangle_{(T,\rho)}$.

Following the analysis of the quark sector, the in-medium modification of the gluon condensate is determined by adopting the results from Ref. \cite{Kumar:2013tna}. Accordingly, its temperature and density-dependent behavior is parameterized as follows:
\begin{align} \label{gluonfit}
\left\langle \frac{\alpha_s}{\pi}G^2\right\rangle_{(T, \rho)}=&\Big[1.9399-0.0410\, (\rho/\rho_{\text{sat}}) -0.0067\,(\rho/\rho_{\text{sat}})^2 + 0.0011\, (\rho/\rho_{\text{sat}})^3 \nonumber\\
&-0.1775\, T +6.9821 \,T^2 - 29.0738\, T^3\Big] 10^{-2} \, \text{GeV}^4.
\end{align}

In addition to the primary condensates discussed above, the following set of light and strange quark operators, along with their mixed representations, are utilized within the hot and dense medium framework:
$\langle q^{\dag} q\rangle_{(T, \rho)} = \frac{3}{2} \rho$, $\langle s^{\dag} s\rangle_{(T, \rho)} = 0$, $\langle \bar{q}g_s\sigma G q\rangle_{(T, \rho)} = m_0^2 \langle \bar{q}q\rangle_{(T, \rho)} + 3\, \text{GeV}^2 \rho$, and $\langle \bar{s}g_s\sigma G s\rangle_{(T, \rho)} = m_0^2 \langle \bar{s}s\rangle_{(T, \rho)} + 3\, y\, \text{GeV}^2\rho$. The higher-order contributions involving the vector-mixed and derivative operators are further defined as $\langle q^{\dag}g_s\sigma G q\rangle_{(T, \rho)} = -0.33 \, \text{GeV}^2 \rho$ and $\langle \bar{q}iD_0iD_0q\rangle_{(T, \rho)} = 0.3\, \text{GeV}^2 \rho - \frac{1}{8}\langle \bar{q}g_s\sigma G q\rangle_{(T, \rho)}$ with analogous expressions for the strange sector scaled by $y = 0.05$. Here, the mass parameter is fixed at $m_0^2 = 0.8 \, \text{GeV}^2$. The current quark masses used in the calculations are taken as $m_u = 2.16\pm 0.04~\mathrm{MeV}$ and $m_s = 93.5\pm0.5~\mathrm{MeV}$ \cite{PDG}.

The derivation of the sum rules for the mass and decay constant is finalized by performing a matching between the QCD and hadronic contributions to the correlation function. Consequently,
\begin{equation}
\label{corre7}
\mathbf{ \hat{B}}\big[p^*_{\mu}\Pi^{QCD}_{\mu\nu}\big] (T,\rho)=\mathbf{\Omega}^*_1(M^2,s^*_0,T,\rho) p_{\nu} + \mathbf{ \Omega}^*_2(M^2,s^*_0,T,\rho) u_{\nu}.
\end{equation}
In this context, $\mathbf{\Omega}^*_{1,2}$ signify the Borel-transformed amplitudes associated with the independent Lorentz structures $p_{\nu}$ and $u_{\nu}$. From this result, the temperature and density dependent sum rules are extracted as follows:
\begin{eqnarray} \label{sumrules}
- f_{PS}^{*2}m_{PS}^{*2} e^{-\mu^2/M^2}&=&\mathbf{\Omega}^*_1(M^2,s^*_0,T,\rho), \nonumber \\
  f_{PS}^{*2}m_{PS}^{*2} \Sigma_{\upsilon}e^{-\mu^2/M^2}&=&\mathbf{\Omega}^*_2(M^2,s^*_0,T,\rho), \nonumber \\
   \Sigma_{\upsilon} &=& - \frac{\mathbf{\Omega}^*_2(M^2,s^*_0,T,\rho)}{\mathbf{\Omega}^*_1(M^2,s^*_0,T,\rho) }.
\end{eqnarray}
The explicit form of the QCD side for the $K^+$ meson sum rules, denoted by $\mathbf{\Omega}^*_{1,2}(M^2,s^*_0,T,\rho)$ in Eq. \eqref{sumrules}, is relegated to Appendix B. Throughout the rest of the article, we adopt the simplified notation $s_0(T,\rho)$ to represent the continuum threshold $s_0^*$, keeping its dependence on temperature and density implicit.


\section{Numerical Analysis of Spectroscopic Parameters}

We perform a numerical analysis of the QCDSRS derived in Eqs. (\ref{sumrules}) to extract the $K^{\pm}$ masses, decay constants, the medium-induced mass shift, and discuss the critical point at finite temperature and density. The calculation incorporates fundamental QCD parameters such as quark masses, and both vacuum and in-medium condensates. 

The QCDSRS in Eqs. (\ref{sumrules}) involve two auxiliary parameters: the Borel mass parameter $M^2$, and the medium-dependent continuum threshold, $s_0 (T,\rho)$. To ensure a reliable application of the sum rule formalism, we impose the standard dual constraints. The upper limit on $M^2$ is fixed by requiring pole dominance, i.e., the ground state contribution must exceed that of excited states and the continuum. The lower limit is set by demanding OPE convergence, requiring the perturbative contribution to dominate  over condensate terms and imposing a convergent hierarchy on the condensate expansion.

We determine the working windows for these parameters following standard QCDSRS criteria, ensuring that predictions for physical observables show minimal dependence on the parameter choice. The continuum threshold is assumed to scale with the in-medium quark condensate as \cite{Dominguez}:
\begin{equation}
\label{eqs0}
s_0(T,\rho) \simeq s_0 \frac{\langle \bar{q}q\rangle_{(T,\rho)}}{\langle \bar{q}q\rangle_0},
\end{equation}
where $s_0$ denotes the vacuum value. Following the stability  analysis, the working windows for the auxiliary parameters are  determined to be:
\begin{equation}\label{eqs1}
M^2 \in [0.4 - 0.6]~\text{GeV}^2, \qquad s_0 \in [0.62 - 0.98]~\text{GeV}^2.
\end{equation}
\begin{figure}[!h]
    \centering
    \begin{minipage}{0.48\textwidth}
        \centering
        \includegraphics[width=\linewidth]{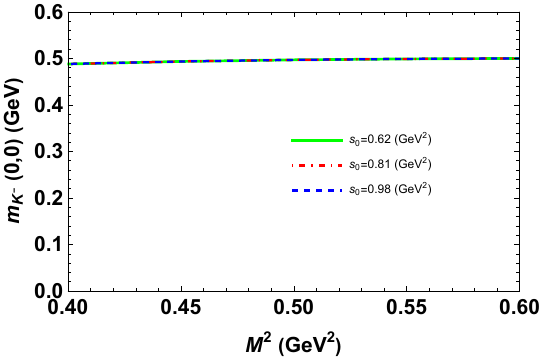}
        \label{fig:left}
    \end{minipage}
    \hfill 
    \begin{minipage}{0.48\textwidth}
        \centering
        \includegraphics[width=\linewidth]{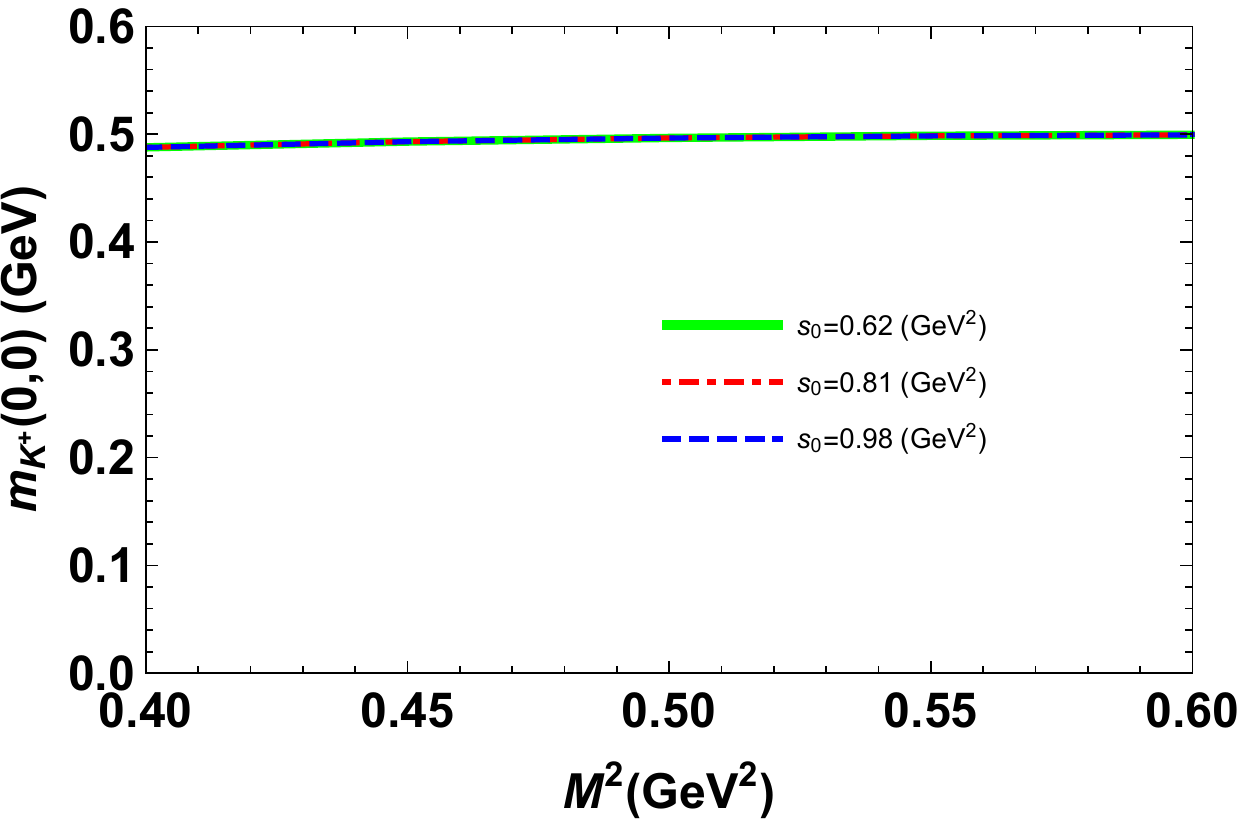}
        \label{fig:right}
    \end{minipage}
    \vspace{-90pt}
    \caption{Dependence of the vacuum masses of $K^{-}$ (left) and $K^{+}$ (right) on the Borel mass parameter $M^2$ for various values of the continuum threshold $s_0$.}
    \label{Fig:1}
\end{figure}

In Fig. (\ref{Fig:1}), we display the $K^-$ and $K^+$ vacuum masses (obtained in the limit  $ T, \rho \to 0$)  as functions of $M^2$ for three representative  values of $s_0$ within the stability window. The masses remain stable across the entire working region, validating the chosen parameter windows. The uncertainties quoted below reflect the residual variation of the extracted observables within these windows.

Employing the average values of $M^2$ and $s_0$, we obtain the following masses and decay constants for the $K^-$ and $K^+$  in vacuum:
\begin{align}
& m_{K^-} = 494.6_{-6.9}^{+4.9}  \, \textrm{MeV}  \, ; \, f_{K^-} =157.3_{-2.9}^{+4.1}  \, \textrm{MeV}, \nonumber \\
& m_{K^+} = 494.5_{-6.8}^{+4.9}  \, \textrm{MeV}  \,; \,  f_{K^+} = 156.9_{-2.8}^{+4.0} \, \textrm{MeV}. 
\end{align}
where the asymmetric uncertainties arise from the variation of $M^2$ and $s_0$ within the windows given in Eq. (\ref{eqs1}). These results 
are in good agreement with the PDG values $m_K = 493.677\pm 0.015$ MeV and $f_K = 155.7\pm 0.3$ MeV \cite{PDG}, at the level of less than 
$0.2\,\%$ and $1.0\,\%$ deviation, respectively. The near-degeneracy of the $K^-$ and $K^+$ masses is consistent with CPT invariance and the small 
isospin-breaking effects in the QCD sum rule framework.

\subsection{Masses and decay constants evolutions in hot and dense medium}
\begin{figure}[!h]
    \centering
    \begin{minipage}{0.48\textwidth}
        \centering
        \includegraphics[width=\linewidth]{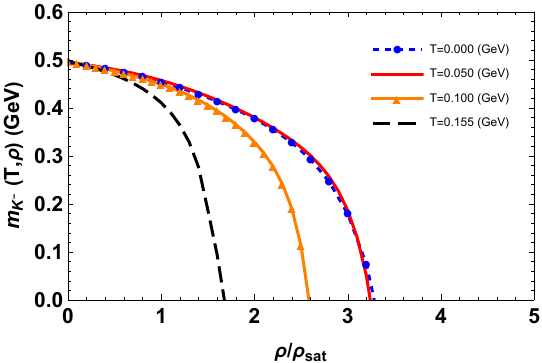}
        \label{fig:left}
    \end{minipage}
    \hfill 
    \begin{minipage}{0.48\textwidth}
        \centering
        \includegraphics[width=\linewidth]{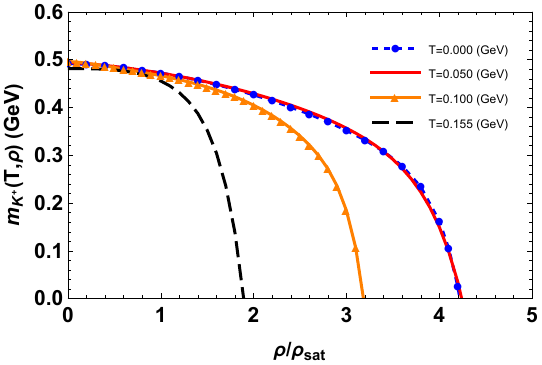}
        \label{fig:right}
    \end{minipage}
    \vspace{-10pt}
    \caption{Density dependence of the modified masses of $K^{-}$ (left panel) and $K^{+}$ (right panel) normalized to the nuclear saturation density ($0\leqslant\rho/\rho_{\text{sat}}\leqslant 5$), at several fixed temperatures ($T=0, 0.050, 0.100, 0.155$ GeV).}
\label{Fig:2}
\end{figure}

Figure (\ref{Fig:2}) shows the density dependence of the effective masses of $K^-$ (left panel) and $K^+$ (right panel) for several temperatures ($T=0, 50, 100,$ and $155$ MeV), with the density normalized to the nuclear saturation density $\rho_{\text{sat}}$. At low temperatures ($T \lesssim 50$ MeV), the effective masses of both mesons exhibits a comparable dependence on nuclear density. As the temperature increases, however, the effective mass profiles of $K^-$ and $K^+$ develop an increasingly pronounced  splitting, in marked contrast to their nearly degenerate behavior in the cold nuclear regime. 

At $T=0$, the effective masses of $K^-$ and $K^+$ reach their vanishing point at approximately $3.23\,\rho_{\textrm{sat}}$ and $4.21\, \rho_{\textrm{sat}}$, respectively. These critical densities are directly relevant to the onset of kaon condensation in dense nuclear matter \cite{KAPLAN198657,KapNelson}. At higher temperatures, the critical point 
decreases markedly: for $K^-$, the effective mass vanishes at $\rho = 2.60 \, \rho_{\textrm{sat}}$ ($T=100$ MeV) and $\rho = 1.63 \, \rho_{\textrm{sat}}$ ($T=155$ MeV). For $K^+$ the corresponding critical densities are $\rho = 3.19 \, \rho_{\textrm{sat}}$  and $\rho = 1.82 \, \rho_{\textrm{sat}}$ at the same temperatures, following a qualitatively similar but systematically shifted pattern relative to $K^-$. This temperature-driven reduction of the critical point reflects the interplay between thermal fluctuations and in-medium condensate suppression.  We have evaluated the relative contributions of the condensates for both the $K^-$ and $K^+$ channels at different densities by computing their average values. For $T=(0-50)$, $100$, and $155$ MeV, the average perturbative contributions are approximately $34.47\%$ ($34.68\%$), $35.19\%$ ($35.37\%$), and $30.87\%$ ($31.09\%$) for the $K^-$ ($K^+$) channel, respectively.  The average quark condensate contributions are approximately $30.72\%$ ($30.92\%$), $29.33\%$ ($29.55\%$), and $31.18\%$ ($38.14\%$), while the corresponding average mixed quark-gluon condensate contributions are about $34.81\%$ ($34.40\%$), $35.48\%$ ($35.08\%$), and $37.95\%$ ($30.77\%$) for the $K^-$ ($K^+$) channel, respectively. These results show that at low temperatures the contributions of the OPE terms are approximately comparable; however, as the temperature increases, the contributions of the condensates become significantly enhanced. In the $K^-$ channel, the mixed quark-gluon condensate becomes the dominant contribution, whereas in the $K^+$ channel the quark condensate becomes increasingly dominant and plays a crucial role in the in-medium behavior. This highlights the different sensitivity of the two channels to the underlying condensate dynamics.
 
\begin{figure}[!h]
    \centering
    \begin{minipage}{0.48\textwidth}
        \centering
        \includegraphics[width=\linewidth]{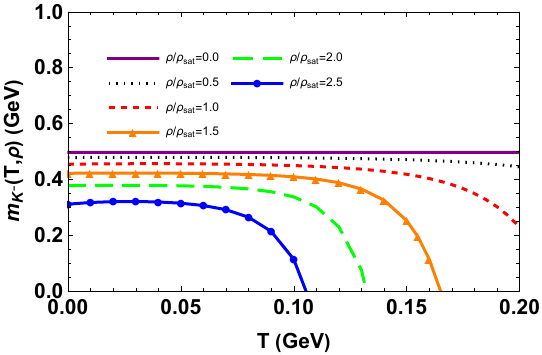}
        \label{fig:left}
    \end{minipage}
    \hfill 
    \begin{minipage}{0.48\textwidth}
        \centering
        \includegraphics[width=\linewidth]{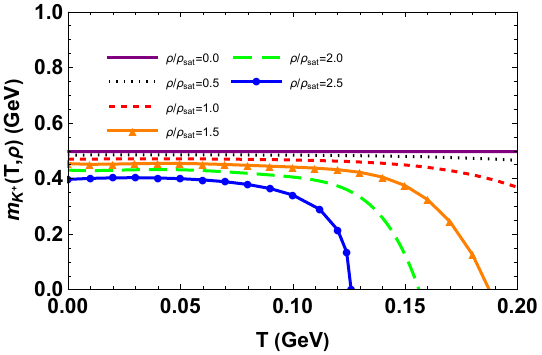}
        \label{fig:right}
    \end{minipage}
    \vspace{-10pt}
    \caption{The modified masses of $K^{-}$ (left panel) and $K^{+}$ (right panel) as a function of temperature for fixed baryon densities ($\rho/\rho_{\text{sat}}=0.0, 0.5, 1.0, 1.5, 2.0., 2.5$) over the range ($0\leqslant T\leqslant 0.20$ GeV).}
\label{Fig:3}
\end{figure}

Figure (\ref{Fig:3}) presents the temperature dependence of the effective masses of $K^-$ (left panel) and $K^+$ (right panel) over the range $0\leqslant T\leqslant 200$ MeV, for fixed baryon densities $\rho/\rho_{\text{sat}}=0.0, 0.5, 1.0, 1.5, 2.0.,$ and $2.5$.  The results are consistent with the density-dependent profiles shown in Fig. (\ref{Fig:2}). Thermal effects become increasingly dominant at higher  densities, leading to a faster reduction of the effective  masses as the temperature approaches the chiral restoration scale. The $K^-$ meson displays a more pronounced sensitivity to thermal variations than $K^+$, particularly in the high-density regime.

The asymmetric behavior between $K^-$ and $K^+$ has a clear physical origin. The $K^+$  meson interacts with the nuclear 
medium primarily through a repulsive vector potential, while  $K^-$ experiences a strong net attractive potential arising  from the combined scalar and vector interactions, partly  mediated by the $\Lambda(1405)$ sub-threshold resonance \cite{Mishra:2004te,Jurgen}. In accordance with various chiral model predictions \cite{Lutz:1992dv,Tolos2002}, this attractive environment accelerates the reduction of the $K^-$ effective mass as the system approaches chiral symmetry restoration. We note that Brown-Rho scaling \cite{Brown1991}  originally formulated for light vector mesons, provides qualitative support for in-medium mass reduction more broadly, though its direct application to kaons should be understood in this qualitative sense.

Beyond the effective masses, the in-medium behavior of the decay constants provides complementary and equally important information. Since the decay constants are intimately linked to the quark condensates $\langle \bar{q} q\rangle$, they serve as direct order parameters for the restoration of chiral symmetry. Their in-medium modifications probe the evolution of the QCD vacuum structure and the weakening of hadronic interactions under extreme conditions of temperature and density.

\begin{figure}[!h]
    \centering
    \begin{minipage}{0.48\textwidth}
        \centering
        \includegraphics[width=\linewidth]{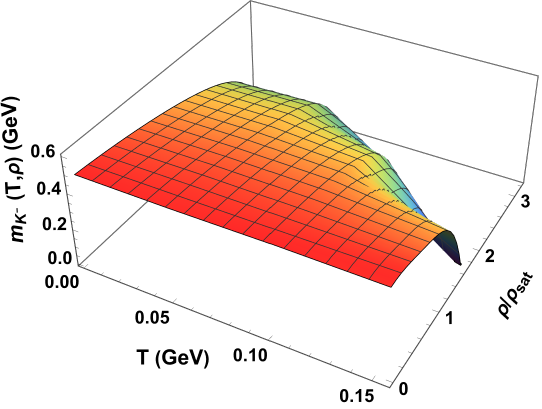}
        \label{fig:left}
    \end{minipage}
    \hfill 
    \begin{minipage}{0.48\textwidth}
        \centering
        \includegraphics[width=\linewidth]{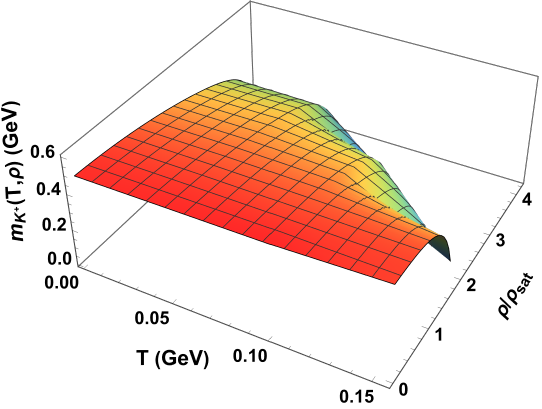}
        \label{fig:right}
    \end{minipage}
    \vspace{-10pt}
    \caption{Density and temperature dependence of the in-medium masses of the $K^{-}$ (left panel) and $K^{+}$ (right panel) mesons shown as three-dimensional surfaces. The horizontal axes represent the baryon density normalized to the nuclear saturation density, $\rho/\rho_{\rm sat}$, and the temperature $T$ (in GeV), while the vertical axis gives the meson mass (in GeV).  The progressive suppression of both masses with increasing density and temperature reflects the partial restoration of chiral symmetry in the hot and dense nuclear medium.}
\label{Fig:33}
\end{figure}

Figure~(\ref{Fig:33}) shows the evolution of the kaon masses in a hot and dense nuclear medium in a three-dimensional representation. For both $K^{-}$ and $K^{+}$ mesons, the masses decrease monotonically with increasing baryon density, reflecting the progressive suppression of the light-quark $\langle \bar{q} q \rangle$ condensate in the medium. Since the masses are directly linked to the order parameter of chiral symmetry breaking, this behavior provides clear evidence for partial restoration of chiral symmetry at finite density.

At zero temperature, the reduction of the masses is moderate up to the nuclear saturation density and becomes more pronounced at higher densities. As the temperature increases, thermal effects significantly enhance this suppression, leading to a faster decrease of both $m_{K^{-}}$ and $m_{K^{+}}$ at a given density. In particular, at $T = 155~\mathrm{MeV}$, the masses approach very small values already at relatively low densities, indicating that thermal fluctuations strongly accelerate the melting of the chiral condensate.

A noticeable splitting between the $K^{-}$ and $K^{+}$ masses emerges at higher densities and temperatures. This asymmetry originates from their different interactions with the nuclear medium: while the $K^{+}$ meson predominantly experiences a repulsive vector interaction, the $K^{-}$ meson is subject to a strong attractive interaction driven by scalar fields and coupled-channel dynamics. Consequently, the mass of the $K^{-}$ meson exhibits a slightly stronger sensitivity to density and temperature compared to that of the $K^{+}$ meson.

Overall, the behavior displayed in Fig.~(\ref{Fig:33}) is consistent with the in-medium mass evolution shown in  Fig.~(\ref{Fig:2}) and  Fig.~(\ref{Fig:3}) supports the interpretation that the modification of kaon decay constants provides a robust signal of chiral symmetry restoration in hot and dense QCD matter.

\begin{figure}[!h]
    \centering
    \begin{minipage}{0.48\textwidth}
        \centering
        \includegraphics[width=\linewidth]{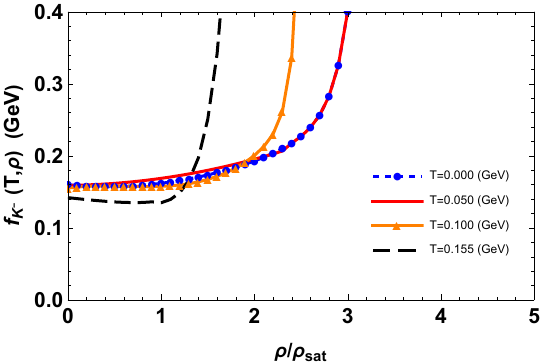}
        \label{fig:left}
    \end{minipage}
    \hfill 
    \begin{minipage}{0.48\textwidth}
        \centering
        \includegraphics[width=\linewidth]{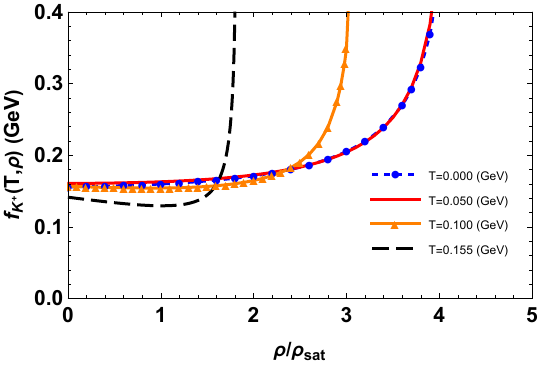}
        \label{fig:right}
    \end{minipage}
    \vspace{-10pt}
    \caption{Density dependence of the in-medium decay constants of $K^{-}$ (left panel) and $K^{+}$ (right panel) normalized to the nuclear saturation density ($\rho_{\text{sat}}$), at several fixed temperatures ($T=0, 0.050, 0.100, 0.155$ GeV).}
\label{Fig:4}
\end{figure}

Figure~(\ref{Fig:4}) presents the baryon-density dependence of the in-medium decay constants $f_{K^-}$ (left panel) and $f_{K^+}$ (right panel) at fixed temperatures $T = 0$, $50$, $100$, and $155$~MeV, with the density normalised to the nuclear saturation density $\rho_{\mathrm{sat}}$. At low to moderate temperatures ($T \lesssim 100$~MeV), both decay constants exhibit qualitatively similar behaviour: they remain nearly flat at low densities and rise rapidly beyond approximately $2\,\rho_{\mathrm{sat}}$ for $K^-$ and $2.5\,\rho_{\mathrm{sat}}$ for $K^+$.  The somewhat lower threshold for $K^-$ reflects its stronger net attractive potential in nuclear matter compared with $K^+$.  At $T = 155$~MeV, both constants first decrease smoothly up to $\rho_{\mathrm{sat}}$ for $K^-$ ($1.5\,\rho_{\mathrm{sat}}$ for $K^+$) before turning upward at higher densities; $K^-$ exhibits this non-monotonic behaviour more markedly than $K^+$.

A qualitatively distinct and physically significant feature emerges at $T = 155$~MeV, near the pseudo-critical temperature $T_c$ for chiral symmetry restoration.  Both $f_{K^-}$ and $f_{K^+}$ display a pronounced non-monotonic dependence on baryon density: an initial suppression at low densities is followed by an anomalous upturn at higher densities.  This structure arises from the competition between two simultaneously active medium effects near $T_c$.  On one hand, the chiral condensate is thermally suppressed, reducing the Goldstone-boson character of the kaon and driving $f_{K^\pm}$ downward.  On the other hand, as the system approaches chiral restoration, the hadronic spectral function undergoes
significant redistribution and the QCD sum-rule analysis begins to receive contributions from continuum states that are absent at lower temperatures.  The resulting effective decay constants therefore encode both the diminishing Goldstone-boson character of the kaon and the growing continuum strength, yielding the observed upturn.  The $K^-$ meson exhibits a more pronounced sensitivity to these effects than $K^+$, consistent with its stronger net attractive potential in the nuclear medium.

It should be emphasised that the upturn seen at $T \approx T_c$ does \emph{not} represent a genuine physical increase of the kaon decay coupling in the conventional sense. Rather, it signals that the sum-rule framework is operating near the boundary of its validity: the standard pole-plus-continuum ansatz for the spectral function becomes increasingly inadequate, and higher-order corrections to the OPE become non-negligible.

\begin{figure}[!h]
    \centering
    \begin{minipage}{0.48\textwidth}
        \centering
        \includegraphics[width=\linewidth]{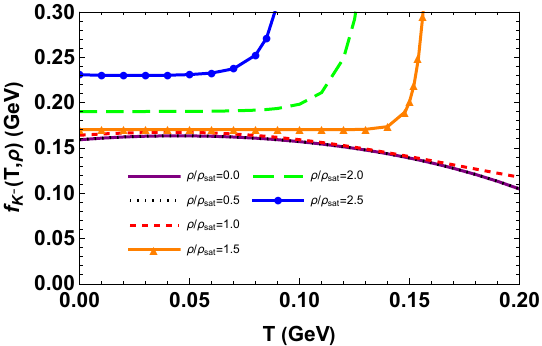}
        \label{fig:left}
    \end{minipage}
    \hfill 
    \begin{minipage}{0.48\textwidth}
        \centering
        \includegraphics[width=\linewidth]{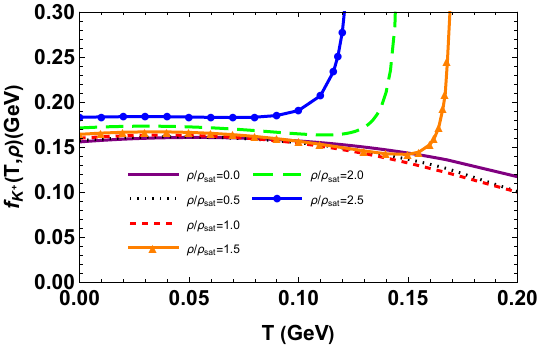}
        \label{fig:right}
    \end{minipage}
    \vspace{-10pt}
    \caption{Temperature dependence of the in-medium decay constants of $K^{-}$ (left panel) and $K^{+}$ (right panel) over the range ($0\leqslant T\leqslant 0.20$ GeV) for fixed baryon densities ($\rho/\rho_{\text{sat}}=0.0, 0.5, 1.0, 1.5, 2.0., 2.5$).}
\label{Fig:5}
\end{figure}

Figure~(\ref{Fig:5}) displays the temperature dependence of $f_{K^-}$ (left panel) and $f_{K^+}$ (right panel) over the range $0 \leq T \leq 200$~MeV for fixed baryon densities $\rho/\rho_{\mathrm{sat}} = 0.0, 0.5, 1.0, 1.5, 2.0$, and $2.5$, confirming and extending the observations from Fig.~(\ref{Fig:4}).  At zero or low baryon density, both decay constants
decrease smoothly and nearly identically with increasing temperature, reflecting the thermal melting of the chiral condensate in a dilute medium.

At high baryon densities ($\rho/\rho_{\mathrm{sat}} \geq 2.0$), a sharp upturn appears at temperatures approaching $T \approx T_c$.  This feature arises from the same mechanism discussed above: the dominance of higher-dimensional OPE condensates and the redistribution of spectral strength toward the continuum
threshold near the chiral crossover.  As stressed in the context of Fig.~(\ref{Fig:4}), this upturn should not be interpreted as a physical
enhancement of the decay constant, but as an indicator that subleading condensate contributions and thermally broadened spectral functions must be incorporated when
analysing kaon properties in the high-$T$, high-$\rho$ regime.  These observations underline the need for improved spectral ansatz and higher-order OPE input in QCD sum-rule studies relevant to heavy-ion collisions and neutron-star merger environments.



\subsection{Mass splitting}
In vacuum, the masses of charged kaons exhibit a nearly perfect symmetry, with only a marginal difference between the $K^-$ and $K^+$ as documented by the PDG \cite{PDG}, arising from electromagnetic corrections within the SM. However, within a hot and dense nuclear medium, such as that created in heavy-ion collisions, this near-degeneracy is profoundly altered. The resulting in-medium mass shift and the subsequent splitting between $K^-$ and $K^+$ are of paramount importance, as they originate from the distinct interactions of strange and anti-strange quarks with the surrounding matter.  As also previously mentioned,  this modification occurs because the $K^{+}$ (containing an anti-strange
quark, $u\bar{s}$) experiences predominantly repulsive interactions, while the $K^{-}$ (containing a strange quark, $\bar{u}s$) feels a strong attractive potential via enhanced $s$-wave couplings with nucleons~\cite{Jurgen,Lutz:1992uw,Lutz:1992dv}.

Studying the evolution of this mass shift from vacuum to finite temperature and density provides a critical, multi-faceted probe of strong interaction physics. Firstly, it offers direct insight into the behavior of chiral symmetry restoration and the relative strengths of the repulsive vector and attractive scalar potentials in dense matter. Secondly, it provides fundamental constraints on the hadronic equation of state and the possible formation of kaon condensation or other exotic phases. Understanding this phenomenon is essential for interpreting key observables in relativistic heavy-ion experiments—such as kaon production yields and collective flow—and serves as a sensitive test for theoretical frameworks like the NJL model or chiral SU(3) dynamics. 
 
\begin{figure}[!h]
    \centering
    \begin{minipage}{0.48\textwidth}
        \centering
        \includegraphics[width=\linewidth]{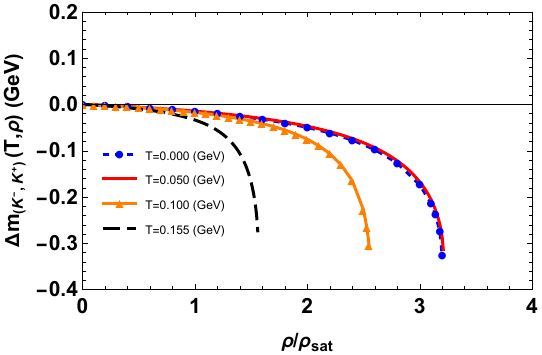}
        \label{fig:left}
    \end{minipage}
    \vspace{-10pt}
    \caption{Evolution of the in-medium mass difference $\Delta m = m_{K^-} - m_{K^+}$ with respect to the nuclear matter density for several fixed temperatures.}\label{Fig:6}
\end{figure}

The density dependence of the mass splitting between $K^-$ and $K^+$ mesons, defined as $\Delta m_{(K^-, K^+)} = m_{K^-} - m_{K^+}$, is illustrated in Fig.~(\ref{Fig:6}) for four distinct temperatures. In the zero-density limit ($\rho = 0$), which corresponds to the vacuum case, the mass difference is nearly zero across all temperatures, consistent with free-space values. However, as the nuclear matter density increases, $\Delta m$ exhibits a significant downward trend across all investigated temperatures, moving into negative values. This behavior indicates that the effective mass of the $K^-$ meson decreases more rapidly than that of the $K^+$ meson in the medium. In the low-temperature regime ($T=0$ and $50$ MeV), representing cold nuclear matter, the mass difference $\Delta m$ remains relatively small at low densities but drops sharply as the density approaches $\rho \approx 3.2 \,\rho_{\text{sat}}$. This pronounced splitting is primarily driven by the interplay between the scalar and vector potentials. While the attractive scalar potential affects both mesons similarly, the vector potential—coupled to the baryon density—acts differently based on the mesons' quantum numbers. Specifically, due to G-parity transformation, the vector potential provides a repulsive shift for the $K^+$ meson but an attractive one for its antiparticle partner, the $K^-$ \cite{Fang}. This opposite coupling, dictated by their distinct quark content
($u\bar{s}$ for $K^{+}$ versus $\bar{u}s$ for $K^{-}$),  effectively amplifies the mass gap, leading to the dramatic downward trend of $\Delta m$ in dense matter.

Conversely, thermal effects significantly modulate this evolution. At elevated temperatures ($T=100$ and $155$ MeV), the splitting curves shift upward toward the zero-line, indicating a reduction in the mass gap compared to the cold regime. This trend suggests a partial restoration of the vacuum symmetry in the hot and dense medium, as increasing thermal fluctuations tend to wash out the distinct scalar and vector potentials experienced by the strange and anti-strange quarks. Notably, at the highest temperature ($T=155$ MeV)—a regime representative of the fireball created in relativistic heavy-ion collisions—the curve terminates near $\rho \approx 1.5 \rho_{\text{sat}}$, likely reflecting the proximity to the chiral phase transition or the limit of the model's stability. Consequently, the in-medium mass splitting $\Delta m$ serves as a multi-faceted probe: it remains most dramatic in cold, dense matter where density effects dominate, yet becomes significantly suppressed under the extreme thermal conditions of a heavy-ion collision.

\begin{figure}[!h]
    \centering
    \begin{minipage}{0.48\textwidth}
        \centering
        \includegraphics[width=\linewidth]{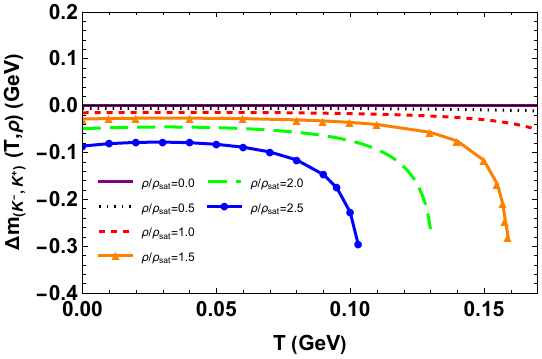}
        \label{fig:left}
    \end{minipage}
    \vspace{-10pt}
    \caption{Temperature dependence of the mass splitting $\Delta m = m_{K^-} - m_{K^+}$ at various constant nuclear matter densities.}\label{Fig:7}
\end{figure}

Complementing the density-dependent analysis, Fig. (\ref{Fig:7}) illustrates the temperature evolution of the $K^\pm$ mass splitting, $\Delta m$, for several fixed nuclear densities. Consistent with Fig. (\ref{Fig:6}), the splitting is negligible at $\rho = 0$ but becomes increasingly negative with rising density.  The mass splitting reaches $|\Delta m|\sim 0.35$~GeV near $\rho\simeq 3.2\,\rho_{\text{sat}}$ at $T=0$ and is partially quenched by thermal fluctuations. In the low-density regime ($\rho \leq 1.0 \rho_{\text{sat}}$), the mass difference remains relatively stable against thermal variations. However, at higher densities ($\rho \geq 1.5 \rho_{\text{sat}}$), a dramatic transition is observed: as the temperature exceeds $T \approx 100$ MeV, $\Delta m$ exhibits a pronounced downward trend. This behavior suggests that in a compressed medium, thermal fluctuations further amplify the disparity between the $s$ and $\bar{s}$ interactions, leading to a significant enhancement of the mass gap.  To better understand this behavior, we analyze the relative contributions of different terms in the OPE for the $K^\pm$ mass splitting. We find that at $T=(0-50)$,  $100$, and $155$ MeV, the quark condensate contributions are approximately $4.55\%$, $9.53\%$, and $32.17\%$, respectively, while the mixed quark-gluon condensate contributions are about $95.45\%$, $90.47\%$, and $67.83\%$. These results clearly demonstrate that the mixed quark-gluon condensate provides the dominant contribution over the temperature range considered. However, the quark condensate contribution increases significantly with temperature and becomes progressively more important. At low temperatures, the mixed quark-gluon condensate dominates, whereas the quark condensate becomes increasingly significant with rising temperature in determining the in-medium behavior.  These results emphasize that the kaon spectral properties are governed by a non-trivial interplay between temperature and density, where thermal excitations act to exacerbate the symmetry breaking already initiated by the baryonic medium.


\subsection{Critical onset density and the QCD phase boundary: temperature dependence and model constraints}

One of the most fundamental and challenging questions in modern nuclear physics is the precise  location and nature of the critical point in the QCD phase diagram—the hypothesized endpoint of the first-order phase transition line separating the hadronic and quark-gluon plasma phases. While first-principles lattice QCD calculations at finite baryon density are hindered by the notorious sign problem, and experimental efforts at facilities such as RHIC-BES, NA61/SHINE, and the forthcoming FAIR and NICA facilities continue to search for definitive signatures~\cite{STAR:2020dav,NA61SHINE:2020czq,CBM:2016kpk}, a comprehensive theoretical understanding remains elusive. Effective model approaches—such as the Nambu--Jona-Lasinio (NJL) model, Polyakov-loop extended quark-meson (PQM) models, and functional renormalization group (FRG) methods—provide valuable complementary insights, but 
each carries intrinsic model dependence and parametric uncertainties~\cite{Buballa:2003qv,Schaefer:2006sr,Schaefer:2009ui}. It is in this context that QCD sum rule analyses, such as the one presented here, play an important complementary role: they provide analytically controlled, model-independent constraints on in-medium hadronic properties directly connected to the behavior of the quark condensate, thereby offering an independent probe of the onset of chiral symmetry restoration.
\begin{figure}[!h]
    \centering
    \begin{minipage}{0.58\textwidth}
        \centering
        \includegraphics[width=\linewidth]{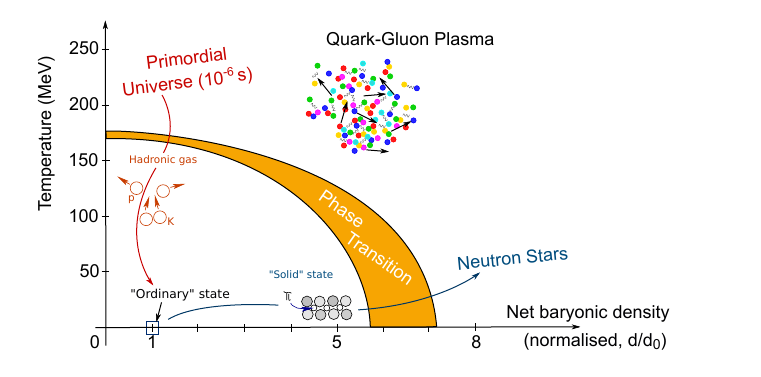}
        \label{fig:left}
    \end{minipage}
    \vspace{-10pt}
    \caption{Schematic QCD phase diagram of hadronic matter adapted from \cite{Maire2011},  illustrating the trajectory from the hadronic phase toward the chiral crossover region.} \label{Fig:9}
\end{figure}

Rather than claiming a precise determination of the QCD critical point—a task that lies beyond the scope of any current effective model—we focus on identifying the region where the hadronic medium begins to undergo significant structural modifications associated with partial chiral symmetry restoration. Specifically, we introduce the quantity $\rho_c$, defined as the density at which the in-medium modifications of kaon properties become pronounced enough to signal the onset of the transition toward the chirally restored phase. This quantity should be interpreted not as a sharp phase boundary, but as an \emph{onset density}—an indicator of the threshold beyond which the assumptions underlying the hadronic description begin to break down and quark degrees of freedom become increasingly relevant. This distinction is physically important: the QCD phase transition at finite temperature and moderate baryon density is expected to be a smooth crossover rather than a discontinuous transition~\cite{Aoki:2006we}, which implies that the notion of a sharply defined critical point is itself an idealization. Accordingly, the values of $\rho_c$ extracted in this work should be understood as delineating the region in the $(T,\rho)$ plane where the hadronic medium enters the vicinity of the chiral crossover.

Our numerical results, summarized inFig.~(\ref{Fig:9}), reveal a pronounced temperature dependence of this characteristic density, consistent with the generic expectations 
from chiral effective models and lattice QCD simulations at finite baryon density~\cite{Aoki:2006we, Bazavov:2017dus}. In the low-temperature regime ($T \lesssim 100$~MeV), relevant for neutron star interiors~\cite{Lattimer:2015nhk} and low-energy heavy-ion collisions, the onset density is found to lie within the range $\rho_c \simeq (1.2-1.4)\,\rho_{\rm sat}$, where $\rho_{\rm sat}$ denotes the nuclear saturation density. This indicates that the hadronic phase remains relatively robust until densities significantly exceeding nuclear saturation are reached, consistent with predictions from chiral perturbation theory and QCD sum rule analyses, which indicate that chiral symmetry breaking remains robust in dense, cold nuclear environments~\cite{Hatsuda:1994pi, Cohen:1994wm}.

However, as the temperature increases into the range $T \sim (100-155)$~MeV—the regime probed by beam energy scan programs and future facilities—a dramatic reduction in $\rho_c$ is observed, with the onset density dropping to approximately $\rho_c \simeq 0.45\,\rho_{\rm sat}$. This significant downward shift reflects the synergistic effect of thermal and density-induced fluctuations in destabilizing the chiral condensate, a phenomenon well-documented in NJL, PQM, and FRG approaches~\cite{Buballa:2003qv,Schaefer:2006sr,Schaefer:2009ui}. More importantly, it suggests that in the hot and dense fireball created in relativistic heavy-ion collisions, the system begins its transition toward a chirally restored phase—characterized by enhanced partonic activity and reduced hadronic binding—at densities well below nuclear saturation. This stands in contrast to cold compressed matter, such as that found in neutron star cores, where chiral symmetry breaking remains robust to much higher densities~\cite{Fukushima:2013rx}.

The shaded band in Fig.~(\ref{Fig:9}) illustrates  a trajectory from the high-density, low-temperature corner of the hadronic phase toward the chiral crossover region at moderate temperatures.  Our results presented above lead to a correlation between the density and temperature that is consistent with this diagram and $ \rho-T $ behaviour. This trajectory should be understood as indicative of the onset of significant medium modifications, 
rather than a precise phase boundary. Indeed, the inherent uncertainties in our approach—stemming from the truncation of the operator product expansion, the simplified spectral function ansatz, and the parametric dependence on input condensates—preclude a precise determination of the critical point itself. Nevertheless, our results provide robust qualitative insights and reinforce the notion that temperature acts as a more efficient driver of chiral restoration than baryon density alone, particularly in the intermediate-density regime accessible to 
current and next-generation experimental programs. This behavior is furthermore consistent with the crossover nature of the QCD phase transition as predicted by lattice QCD extrapolations and phenomenological studies of the QCD phase diagram~\cite{Ding:2015ona, Borsanyi:2020fev}.

In summary, while the exact location of the QCD critical point remains an open question, studies such as the present one contribute to delineating the boundaries of the hadronic phase 
and identifying the conditions under which chiral symmetry restoration begins. By providing a systematic, QCD-founded analysis of in-medium kaon modifications, our results offer valuable input for interpreting experimental data and for guiding the development of more sophisticated theoretical frameworks capable of eventually pinning down the critical point with greater precision.


\section{Summary and Concluding remarks}

In this work, we have presented a comprehensive and systematic analysis of the in-medium modifications of charged kaons (\(K^{\pm}\)) within the framework of QCDSRS, with a particular focus on their behavior across the temperature (\(T\)) and baryon density (\(\rho\)) plane. By incorporating the fundamental temperature and density dependencies of the key QCD condensates—namely the chiral condensate \(\langle \bar{q} q\rangle\), the strange quark condensate \(\langle \bar{s} s\rangle\), and the gluon condensate \(\langle \frac{\alpha_s}{\pi} G^2\rangle\)—into the OPE, we have derived sum rules that reliably connect the underlying quark-gluon dynamics to the hadronic observables of interest. This approach has enabled us to quantify the effects of partial chiral symmetry restoration on the effective masses \(m_{K^{\pm}}\) and decay constants \(f_{K^{\pm}}\) in a hot and dense nuclear medium.

Our findings reveal a rich and nuanced picture of kaon properties under extreme conditions. A central result is the confirmation and systematic quantification of the pronounced mass splitting \(\Delta m = m_{K^-} - m_{K^+}\) in dense baryonic matter. At zero and low temperatures, this splitting becomes increasingly negative with density, driven by the opposite sign of the vector interaction for \(K^+\) (repulsive) and \(K^-\) (attractive) in accordance with the Weinberg-Tomozawa term. This asymmetry, rooted in their distinct quark content (\(u\bar{s}\) vs. \(\bar{u}s\)), leads to a significantly faster drop in the \(K^-\) effective mass compared to its particle partner. At higher temperatures, however, thermal fluctuations partially counteract this effect, shifting the splitting curves toward zero and indicating a gradual restoration of vacuum symmetry.

The in-medium behavior of the decay constants \(f_{K^{\pm}}\) provides an equally critical probe of chiral restoration. A particularly intriguing feature emerges near the pseudo-critical temperature \(T_c \sim 155\) MeV, where both \(f_{K^-}\) and \(f_{K^+}\) exhibit a non-monotonic dependence on density—an initial suppression followed by an upturn. We have interpreted this not as a physical enhancement of the decay constant, but as a signature of the limitations of the standard QCDSRS pole-plus-continuum ansatz near the phase boundary. This upturn signals the increasing importance of higher-order condensates and the redistribution of spectral strength, underscoring the need for refined spectral functions in future analyses of the high-\(T\), high-\(\rho\) regime.

Among the key outcomes of this study is the extraction of the critical onset density  $\rho_c$, which we have identified as the threshold beyond which the hadronic medium begins to undergo significant structural modifications associated with partial chiral symmetry restoration. As emphasized throughout this work, $\rho_c$ should not be interpreted as a sharply defined phase boundary or a precise determination of the QCD critical point, but rather as an indicator of the density regime where hadronic descriptions begin to break down and quark degrees of freedom become increasingly relevant. With this interpretation in mind, our results demonstrate a pronounced temperature dependence of this threshold. In the cold, dense 
matter regime ($T \lesssim 100$~MeV), relevant to neutron star interiors and low-energy heavy-ion collisions, the onset density resides at supra-saturation values 
$\rho_c \simeq (1.2-1.4)\,\rho_{\rm sat}$, reflecting the robustness of the hadronic phase under cold compression. However, as the temperature rises into the range accessible by beam 
energy scan programs and future FAIR/NICA facilities $T \sim (100-155)$~MeV, $\rho_c$ drops dramatically to sub-saturation values, $\rho_c \simeq 0.45\,\rho_{\rm sat}$. This finding 
reinforces the notion that temperature acts as a more efficient driver of chiral restoration than baryon density alone: the hot and dense fireball created in relativistic heavy-ion 
collisions enters the chiral crossover region at considerably lower densities than the cold, compressed matter found in astrophysical compact objects, suggesting a qualitatively different pathway toward chiral symmetry restoration in the two physical environments.

In conclusion, this work provides a unified and quantitative picture of strangeness dynamics in hot and dense QCD matter. Our systematic QCDSR analysis yields several key insights:
\begin{itemize}
    \item \textbf{Flavor-Dependent Medium Effects:} The in-medium behavior of \(K^+\) and \(K^-\) is fundamentally asymmetric due to their quark content, leading to a density-driven mass splitting that serves as a clear signal of the changing vacuum structure.
    \item \textbf{Chiral Order Parameters:} The decay constants \(f_{K^{\pm}}\) directly track the in-medium evolution of the chiral condensate, with their suppression providing robust evidence for partial chiral symmetry restoration.
   \item \textbf{Temperature as the Primary Driver:} The critical onset density $\rho_c$, beyond which significant medium modifications signal the approach to chiral symmetry restoration, drops sharply with increasing temperature. This establishes thermal effects as the dominant mechanism driving the system toward the chiral crossover in the regime probed by current and future experiments, and demonstrates that temperature is a more efficient control parameter than baryon density alone in this intermediate-density regime.
   \item \textbf{Temperature-Density Correlation:} The obtained results and behaviors are consistent with the QCD phase diagram.
    \item \textbf{Bridging Theory and Experiment:} Our calculated results for \(m_{K^{\pm}}(T,\rho)\) and \(f_{K^{\pm}}(T,\rho)\) offer essential, QCD-based input for interpreting experimental data from HADES, STAR, NA61/SHINE, and the upcoming CBM experiment at FAIR. They are also crucial for constraining the hadronic equation of state in astrophysical contexts, particularly regarding the possibility of kaon condensation in neutron stars.
\end{itemize}

Our study firmly establishes that the in-medium modifications of kaons are a sensitive and multi-faceted probe of the QCD phase diagram. The results reinforce the crossover nature of the transition at moderate densities and highlight the synergistic role of temperature and density in melting the chiral condensate. Future work should focus on refining the spectral function ansatz within the sum rule approach, incorporating a more rigorous treatment of finite-width effects and higher-order condensate contributions, particularly in the vicinity of the phase boundary. This will further enhance the precision of QCDSR predictions and solidify their role as a cornerstone tool in the exploration of strongly interacting matter under extreme conditions.

\appendix
\section*{Appendix A: The Fermi-Dirac Distribution and the Thermal Propagator}
In this section, we provide a brief derivation of the Fermi-Dirac distribution function as it appears within the real-time formalism of finite-temperature field theory. The perturbative part of the thermal fermion propagator in momentum space is expressed as:
\begin{equation}
\label{ThermalProp}
\mathcal{G}_{\textrm{pert}}^{mn} (p)=i (\slashed{p} +m) \Big[\frac{1}{p^2-m^2+i\epsilon}+2\pi i \mathcal{F} (|p_0|)\delta(p^2-m^2)\Big], 
\end{equation}
where $ \mathcal{F} (|p_0|)=1/(e^{\beta|p_0|}+1)$ denotes the Fermi-Dirac distribution function, with $\beta = 1/T$ being the inverse temperature. To obtain the coordinate space representation, we perform a Fourier transformation of Eq.~(\ref{ThermalProp}):
\begin{equation}
\label{ }
\mathcal{G}_{\textrm{pert}}^{mn} (x)= \int\frac{d^4 p}{(2\pi)^4} e^{-ip\cdot x} \mathcal{G}_{\textrm{pert}}^{mn} (p).
\end{equation}
The integration is carried out by decomposing the propagator into its vacuum and thermal components. By employing the Cutkosky rules (or the Sokhotski–Plemelj theorem) to handle the singularity in the propagator, specifically:
\begin{equation}
\label{Cutkosky }
 \left[ \frac{1}{p^2 - m^2 + i\epsilon} \right] = -2\pi i \delta(p^2 - m^2),
\end{equation}
one can isolate the on-shell contributions. After performing the integration over the four-momentum $p$, the perturbative part of the thermal  propagator in coordinate space, which is essential for evaluating the self-energy loops at finite temperature and density, is obtained as
\begin{equation}
\label{ }
\mathcal{G}_{\textrm{pert}}^{mn} (x)=  \Big[1-\frac{1}{e^{\beta/ |x_0|}+1}\Big]\Big(\frac{i}{2\pi^2}\frac{\slashed{x}}{ x^4}-\frac{m_q}{4\pi^2x^2}+i\frac{m_q^{2} }{8\pi^2 } \frac{\slashed{x}}{x^2}\Big)\delta^{mn}.
\end{equation}

\appendix
\section*{Appendix B: Analytic Expressions for the QCD Spectral Density $\mathbf{\Omega}^*_{1,2}$}
In this appendix, we present the functions $\mathbf{\Omega}^*_1(M^2,s^*_0,T,\rho)$ and  $\mathbf{\Omega}^*_2(M^2,s^*_0,T,\rho)$ for the $K^+$ meson, which incorporates temperature and density dependencies along with the physical parameters originating from the hadronic representation of the correlation function, as follows:
\begin{eqnarray}\label{sigma1star}
\mathbf{\Omega}^*_1(M^2,s^*_0,T,\rho)&=&\frac{3}{8\pi^2} (1-\mathcal{F})^2 (m_u+m_s)^2 M^2  \big[e^{-(m_u+m_s)^2/M^2}-e^{s^*_0/M^{2}}\big] -\frac{2 }{3 \text{M}^2}  (1-\mathcal{F})  \Big\{ \left( M^2+2 p_0^2 \right) m_s \langle \bar u g_s \sigma G u \rangle \nonumber \\
&+&  \left( M^2+2 p_0^2 \right) m_u \langle \bar s g_s \sigma G s \rangle -2m_u  \Big [ 4\left(M^2 +2p_0^2\right) \langle \bar s iD_{0} iD_{0}s\rangle -3M^2 \big(M^2  \langle \bar s s\rangle - 2 m_q p_0  \langle s^{\dagger}s\rangle \big)\Big ]  \nonumber \\ 
&+& 2m_s  \Big [- 4\left(M^2 +2p_0^2\right) \langle \bar u i D_{0}iD_{0}u\rangle + 3M^2 \big(M^2  \langle \bar u u\rangle - 2 m_q p_0  \langle u^{\dagger}u\rangle \big) \Big ]\Big\}  \nonumber \\ 
&+& \frac{p_0}{3M^2}\left( \langle u^{\dagger}  g_s \sigma G u\rangle -  \langle s^{\dagger}  g_s \sigma G s\rangle \right),  \nonumber \\ 
\mathbf{\Omega}^*_2(M^2,s^*_0,T,\rho)&=&\frac{1}{6} \big(\langle u^\dagger g_s \sigma G u\rangle-\langle s^\dagger g_s \sigma G s\rangle \big).
\end{eqnarray}
Note that the  four-dimensional gluon condensate $ \sim \langle G^2\rangle $ gives zero contribution to the results.

\section*{ACKNOWLEDGMENTS} 
K. Azizi thanks Iran national science foundation (INSF) for the partial financial support provided under the elites Grant No. 40405095.


\end{document}